\documentclass[12pt]{article}

\def\thebibliography#1{\section*{References}\list
 {}{\setlength\labelwidth{1.4em}\leftmargin\labelwidth
 \setlength\parsep{0pt}\setlength\itemsep{0pt}
 \setlength{\itemindent}{-\leftmargin}
 \usecounter{enumi}}}

\begin{document}
\begin{center}
{\bf Giant Radio Sources in View of the Dynamical Evolution of FRII-type
Population. II. The Evolutionary Tracks on the $P-D$ and $u_{\rm c}-E_{\rm tot}$
Planes}
\end{center}

\begin{center}
by

{J. Machalski, K.T. Chy\.{z}y \& M. Jamrozy}
\end{center}

\begin{center}
Astronomical Observatory, Jagellonian University, ul. Orla 171,\\ 30--244 Cracow, Poland \\
e--mail: machalsk@oa.uj.edu.pl jamrozy@oa.uj.edu.pl
\end{center}

\begin{abstract}

The time evolution of `fiducial' radio sources derived from fitting the dynamical
model of Kaiser et al. (1997) is compared with the observational data for the
`clan' sources found in the sample of {\sl giant} and normal-size FRII-type
sources published Paper I (Machalski et al. 2004). Each `clan' comprises 3, 4 or
5 sample sources having similar values of the two basic physical parameters: the
jet power $Q_{0}$ and central density of the galaxy nucleus $\rho_{0}$ (determined
in Paper I) but different ages, radio luminosities and axial ratios. These sources
are considered as the `same' source observed at different epochs of its lifetime
and used to fit the evolutionary luminosity--size ($P-D$) and energy
density--total energy ($u_{\rm c}-E_{\rm tot}$) tracks derived from the model
for a `fiducial'
source with $Q_{0}$ and $\rho_{0}$ equal to the means of relevant values obtained
for the `clan' members, as well as to constrain the evolutionary model of the
source dynamics used. In the result we find that (i) The best fit is achieved when
the Kaiser et al.'s model is modified by allowing an evolution of the sources'
cocoon axial ratio with time as suggested by Blundell et al. (1999). (ii) A slow
acceleration of the average expansion speed of the cocoon along the jet axis is
suggested by the `clan' sources. We argue that this acceleration, although minor,
may be real and some supporting arguments come from the well known hydrodynamical
considerations.


\end{abstract}

\section{Introduction}

There are several analytical models of the dynamical evolution of extended powerful
radio sources whose prototype is the radio galaxy Cyg\,A (3C405). All of them are
based on the hydrodynamical expansion of their lobes (or cocoon) caused by the
interaction of light supersonic jets with the ambient material swept-up within the
bow-shock arising behind the jet head, and satisfactory describe the sources'
dynamics. However, only two of them, the most sophisticated ones published by
Kaiser, Dennett-Thorpe and Alexander (1997) [hereafter KDA] and Blundell, Rawlings
and Willott (1999) [hereafter BRW], deal with the sources' energetics and the radio
luminosity evolution with time. These two models differ in one assumption: the KDA
model sustain the Falle's (1991) requirement of a homologous (self-similar) bow-shock
expansion, while in the BRW one the jet-head pressure does not scale with
time in the same way as cocoon pressure, hence it does not expand self-similarly.
As the result, the predicted decrease of the radio luminosity of matured sources
is faster in the BRW model than in the KDA one. Besides, the self-similar bow-shock
expansion would be possible only if the internal source (cocoon) pressure was
always much higher than the external environment pressure, what has been
questioned by X-ray observations of the medium surrounding selected FRII-type
galaxies and quasars (e.g. Hardcastle and Worrall 2000). Therefore {\sl giant} radio
sources with their largest linear sizes and low luminosity are excellent objects
for an observational constraint of the both models predictions.

In Machalski et al. (2004) [hereafter referred to as Paper I] we presented the 
sample of 18 {\sl giant} lobe-dominated radio sources (with projected linear size
$D>1$ Mpc if $H_{0}$=50 km\,s$^{-1}$Mpc$^{-1}$ and $\Omega_{\rm M}$=1) and 54
normal-size sources selected to fulfill the following criteria: (i) the sources
have the FRII-type (Fanaroff and Riley 1974) morphology, (ii) the existing radio
maps allow a suitable determination of their lateral size transversal to the source's
axis which is necessary to specify its (so called) `axial ratio' and volume, and
(iii) their spectral age or the expansion speed, reliably determined with the same
model of the energy losses, are available from the literature.

In Paper I, applying the KDA model chosen for its simplicity in comparison to the
BRW, for each member of the
sample we have determined their basic physical parameters, i.e. the jet power
$Q_{0}$, the central density of the galaxy nucleus $\rho_{0}$, the energy
density and pressure in the lobes/cocoon ($u_{c}$ and $p_{c}$), and the total
energy of the source $E_{\rm tot}$. All these parameters are derived from fitting
the model free parameters to the source redshift, monochromatic radio luminosity,
projected size, and axial ratio, i.e. its observed parameters. Next, these model
and observed parameters have been used to search for
conditions or circumstances under which extragalactic radio sources can reach
their extremal linear extent even of a few Mpc, to check the values of the
internal pressure in their lobes and to compare them to the suggested external
pressure in the sources' environment, etc.

As a result, we argued that (i) the {\sl giant}-size sources do not form a unique
class of extragalactic objects; they are old sources with the jet power
$Q_{0}>3\cdot10^{37}$ W evolved in a relatively low-density environment with
$\rho_{0}<10^{-23}$ kg\,m$^{-3}$, (ii) an apparent increase of the lowest internal
pressure value with redshift, found in the largest sources, is obscured by the
intrinsic dependence of their size on age and the age on redshift, (iii) the ratio
of the jet-head pressure expanding the source cocoon in the jet direction and the
cocoon lateral pressure ($p_{\rm h}/p_{\rm c}$) governing its axial ratio is
dependent on the source's age (as well as on the jet power) which violates the
KDA model assumption of self-similar expansion of the cocoon. We showed that a
departure from self-similarity for large and old sources is justified by the
intrinsic correlation between their observed axial ratios and ages. This forms
an independent observational support for the competitive BRW model.

Moreover, we have realized that for a number of the sample sources the derived
values of their physical parameters $Q_{0}$ and $\rho_{0}$ were very close,
while their ages, luminosities, and linear sizes were significantly different.
This immediately raises a question whether a proximity of the above two fundamental
model parameters appears by chance or it has a real physical meaning. If the
latter was true, observed luminosity and size of these sources should follow
more or less the model time evolution predictions. To check this we assume that
they may be considered as `the same' source observed at different epochs of its
lifetime. Hereafter we call these sources as a `clan' although the members of
each `clan' have rather different redshifts. The above may arise a doubt about
the density $\rho_{0}$ being constant in the KDA model during the source's
lifetime at a given redshift. However, there are no evidences for an evolution of
galaxies' density and density profile during the life of \underline{a radio
source}, thus the neutral assumption that such parameters are independent of
redshift is commonly accepted (cf. Gopal-Krishna and Wiita 2001). Nevertheless,
the galaxy halo can presumably be scaled with the mean matter-density evolution
of the Universe proportional to $(1+z)^{3}$. To check how this effect can influence
the source's parameters derived from the KDA model, we have selected two giants
with the largest and smallest redshift in the sample, i.e. MRC0437-244 and
PKS0319-454 with redshifts of 0.84 and 0.0633, respectively (cf. Table~1 of
Paper~I). Using the cosmological constants adopted in this paper, both sources
have been shifted in the look-back time by their given ages (19 and 180 Myr,
respectively; Table~2 of Paper~I). Then, we have computed the redshift at the
birth-time of these giants, and recalculate their model parameters for such
higher redshifts. The resultant parameters differ only very little from the
original ones (by less than 1\%) justifying the assumption that the density
profile of a galaxy does not evolve during rather short lifetime of radio 
sources in comparison to a time-scale of the matter large structures' evolution. 

Hereafter we
show that these `clans' appear crucial for the observational constraint of the
KDA model predictions of the source's time evolution presented in this paper.
In Section~2 the members of the three `clans' are specified, and the evolutionary
tracks of their model representation on the radio luminosity--linear size ($P-D$)
and energy density--total energy ($u_{\rm c}-E_{\rm tot}$) planes are calculated
and compared the observational data. The results obtained, especially the
necessity of a departure from self-similarity of the lobe (cocoon) expansion in
time, are discussed in Section~3. The conclusions are given in Section~4.

\section{Evolutionary Tracks of Sources}

All dynamical models  enable one to calculate evolutionary tracks of
time-dependent secondary parameters of radio sources if their primary
(time-independent) parameters are given. In the papers of KDA and BRW
the tracks of radio luminosity $P$ versus linear size $D$ were derived
for imaginary sources with assumed values of $Q_{0}$, $\rho_{0}$, $a_{0}$,
$\beta$, and $z$.

In our approach we are able to calculate such evolutionary tracks for actual
sources. The sample of radio sources used in this analysis (cf. Paper I) consists
of the four sets of sources: {\sl giants} (with $D>$1 Mpc), normal-size
high-redshift and low-redshift (with $z\geq$0.5 and $z<$0.5, respectively), and
low-luminosity (with $P_{\rm 1.4}<10^{24.4}$ W\,Hz$^{-1}$sr$^{-1}$). In Section~1
the `clans' of sources have been introduced, i.e.
the sample sources with very close values of the two fundamental parameters
in the KDA model, the jet power $Q_{0}$ and central core density $\rho_{0}$, and
evidently different ages, luminosities, and axial ratios. Since this dynamical
model assumes constant jet power during a source lifetime, and the nucleus density
$\rho_{0}$ is {\sl a priori} constant, members of such a clan can be considered as
`the same' source observed at a number of different epochs throughout its life.
The observed parameters of these members can verify predictions of the model.

\subsection[section]{Clans}

We found 6 clans consisting of 3, 4 or 5 sample sources fulfilling the adopted
selection criterion, namely that the fitted values of $Q_{0}$ and $\rho_{0}$ do
not differ from their mean value by more than 30 \%, i.e. the standard
deviation from the mean is less than 0.114 in the logarithmic scale. Also the
redshift of members should be comparable, however in our limited sample we have
accepted  redshift ratios up to about 3. Leaving out three-member clans, the
remaining three of six clans are marked in Fig.~1a and b, and their members are
listed in Table~1.

\begin{figure}
\vspace{170mm}
\includegraphics{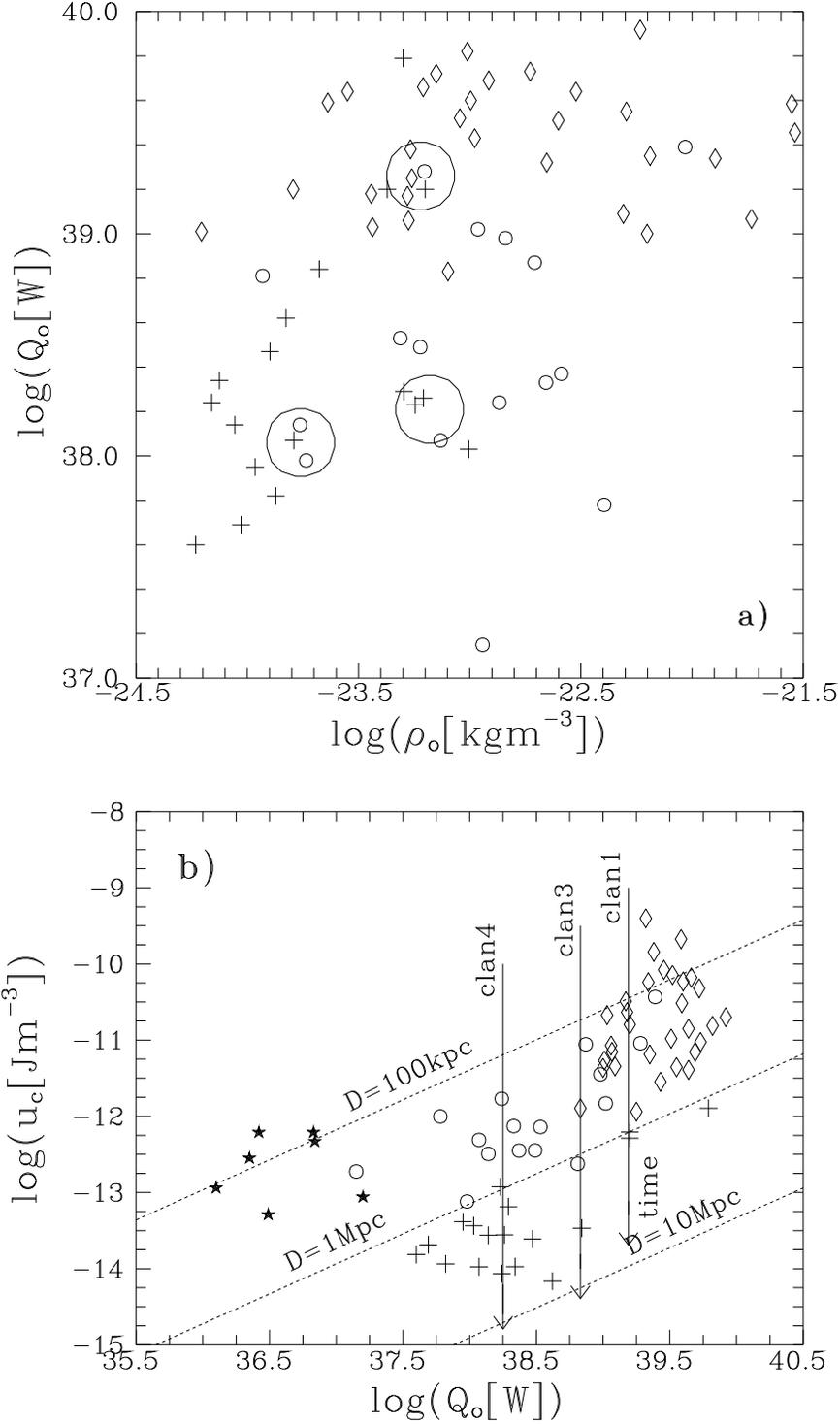}
\caption{{\bf a)} Plots of the jet power $Q_{0}$ against central density of the
core $\rho_{0}$. The {\sl giants} are indicated by crosses, high-redshift sources
-- diamonds, low-redshift sources -- open circles, and low-luminosity sources --
stars. The `clans' of a few sources with comparable values of $Q_{0}$ and
$\rho_{0}$ are marked by large circles with the radius being equal to the standard
deviation of log$Q_{0}$ and log$\rho_{0}$ values for given `clan' members;
{\bf b)} The same `clans' are localized on the $u_{\rm c}$ -- $Q_{0}$ plane by
time axes.}
\end{figure}

\begin{table}[htb]
\begin{center}
\caption{Members of the clans}
\begin{tabular}{@{}llll}
\hline
               & Clan1   & Clan3   & Clan4\\
\hline
$\langle\lg Q_{0}$[W]$\rangle$ & 39.26$\pm$0.08 & 38.84$\pm$0.03 & 38.21$\pm$0.10\\
$\langle\lg\rho_{0}$[kg/m$^{3}]\rangle$ & $-$23.24$\pm$0.04 & $-$23.35$\pm$0.05 & $-$23.22$\pm$0.07\\
$\langle z\rangle$ & 0.72$\pm$0.26   & 0.43$\pm$0.15   & 0.20$\pm$0.09\\
clan members        & 3C263.1 & 3C166   & 3C319\\
               & 3C289   & 3C334   & 1025$-$229\\
               & 3C411   & 1012+488& 0136 +396\\
               & 3C272   & 0821+695& 3C326\\
               & 3C274.1\\
\hline
\end{tabular}
\end{center}
\end{table}

\noindent
The mean values of log$Q_{0}$, log$\rho_{0}$, and redshift in each clan are given
with the standard deviations of their individual values from the mean. 
The members are ordered according to their increasing age. In all three clans
that increase of age is accompanied with an increase of the size and total
energy of the members, altogether with a decrease of their luminosity and energy
density. This qualitative observational behaviour, concordant with the model
expectations, is quantitatively presented in Section~2.2.

\subsection[section]{Time Evolution of the Clans}

In order to check whether observed parameters of the members of a given clan are
consistent with the time evolution derived from the KDA model, a fiducial source
has been created for each clan. This fiducial
source has $Q_{0}$ and $\rho_{0}$ equal to the mean value of these parameters
in a given clan and the evolving axial ratio of the cocoon
$AR(t,Q_{0} )\propto t^{0.23\pm 0.03}Q_{0}^{0.12\pm 0.02}$ [Eq.\,(11) in Paper I).
For each of the clans, a time evolution of its size $D(t)$, luminosity
$P(t)$, energy density $u_{\rm c}(t)$, and total energy $E_{\rm tot}(t)$
have been calculated. The predicted $D(t)$ tracks for the three clans are shown
in Fig.~2a with the dotted lines. The solid lines connect members
of each clan drawn with different symbols. The data points are shown with error
bars (the errors in size are usually smaller than the vertical size of symbols).
The dashed line indicates $D(t)$ for the constant advance speed of 0.1$c$. 
The higher slopes of the modelled tracks with respect to this constant speed seem
to indicate a slowly accelerating expansion of the cocoon in the clans
investigated, however the effect is of low significance because of uncertainties
of the age. More extensive discussion of the above effect is given in 
Section~2.3.

\begin{figure}
\vspace{155mm}
\includegraphics{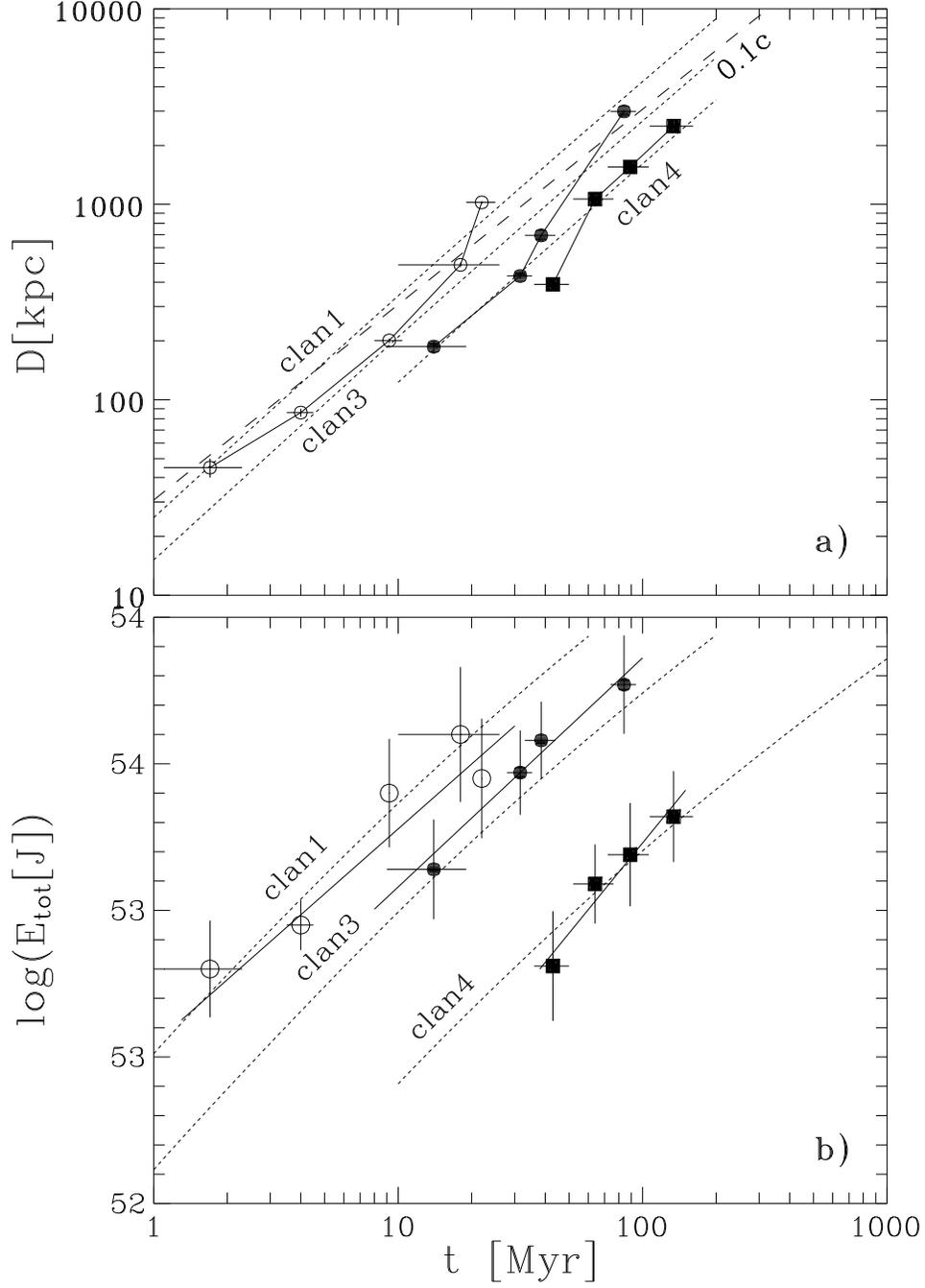}
\caption[]{{\bf a)} Time evolution of the size of the fiducial source for each of
the three observed clans (dotted lines). The dashed straight line indicates a constant
advance speed of 0.1$c$; {\bf b)} Evolution of the total energy of a fiducial
source for each of the clans. The dotted lines show the model prediction, the
shorter solid lines -- the best power-law fit to the data points.}
\end{figure}

The time evolution of the total energy predicted with the KDA model for the three
clans is shown in Figure~2b with the dotted lines. The short solid lines indicate
the best power-law fit to the data points. A correspondence between the model
$E_{\rm tot}(t)$ and the data points is satisfactory.

\subsection[section]{Tracks log$P$--log$D$}

The evolutionary tracks of the fiducial source for the three clans are shown in
Figure~3. The members of separate clans are marked by different symbols as in 
Figure~2a and 2b. The actual age of each source is indicated by the number in a
vicinity of each symbol. The
dashed curves show the tracks calculated from the original KDA model, i.e. with
a constant $AR$ taken as the mean of axial ratios in a given clan. It is
clearly seen that these fits are unsatisfactory. Therefore, we make the
second run of the fits with the KDA model but modified by substitution of 
$R_{\rm T}$ in the formula of Kaiser (2000)
\[{\cal P}_{\rm hc}=(2.14-0.52\beta)R_{\rm T}^{2.04-0.25\beta},\]

\noindent
used in Paper I, by our empirical formula for the age-dependent
$R_{\rm T}(t)\equiv AR/2\propto t^{0.24}$ (cf. Paper I). Hence 

\[{\cal P}_{\rm hc}=0.0056\, t^{0.4},\]

\noindent
where $t$ is expressed in [yr]. This is consistent with the formalism used in the
BRW model in which the injection spectral index is governed by the breaks in the
energy distribution of the particles injected into the cocoon from the hotspot
instead of the constant index of $p=2.14$ in the KDA model. The modified tracks,
marked with the solid curves, show that the evolving $AR$ much better fits the
observed changes of $P$ and $D$. The markers of the same age on these tracks are
connected with a dotted line. Further discussion of the best fit tracks is given
in Section~3.1.

\begin{figure}[t]
\vspace{85mm}
\includegraphics{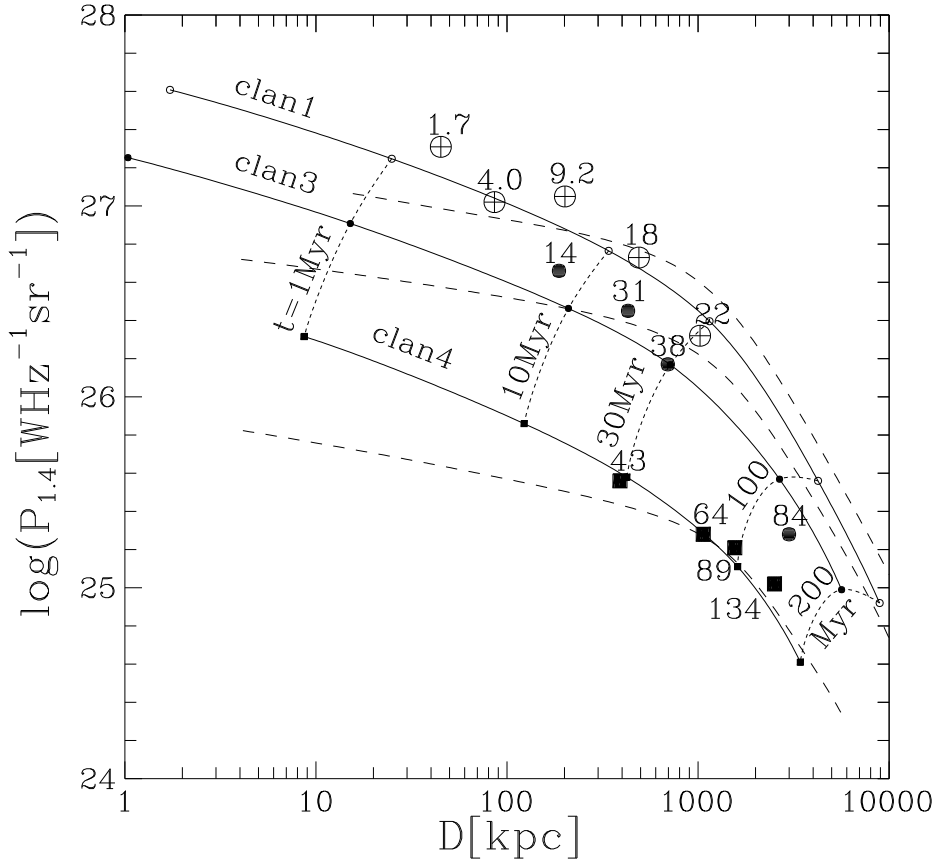}
\caption[]{Evolutionary $P$--$D$ tracks fitted for three clans of sources with
evolving axial ratio $AR$ (solid curves). The markers of the same predicted age 
on each curve are connected with dotted lines. The members of each clan are 
marked with different symbols as those in Figures 2a and 2b. Their actual
age is indicated by a number behind the symbol. The dashed curves indicate
relevant tracks but calculate with a constant $AR$, as in original KDA model.}
\end{figure}

\subsection[section]{Tracks log($u_{\rm eq}$)--log($E_{\rm tot}$)}

The model also allows one to predict the evolution of a source on the energy
density--total energy plane. These tracks are shown in Figure~4 with dashed
curves calculated with the original KDA model and the solid ones with the
modified model.

\begin{figure}
\vspace{83mm}
\includegraphics{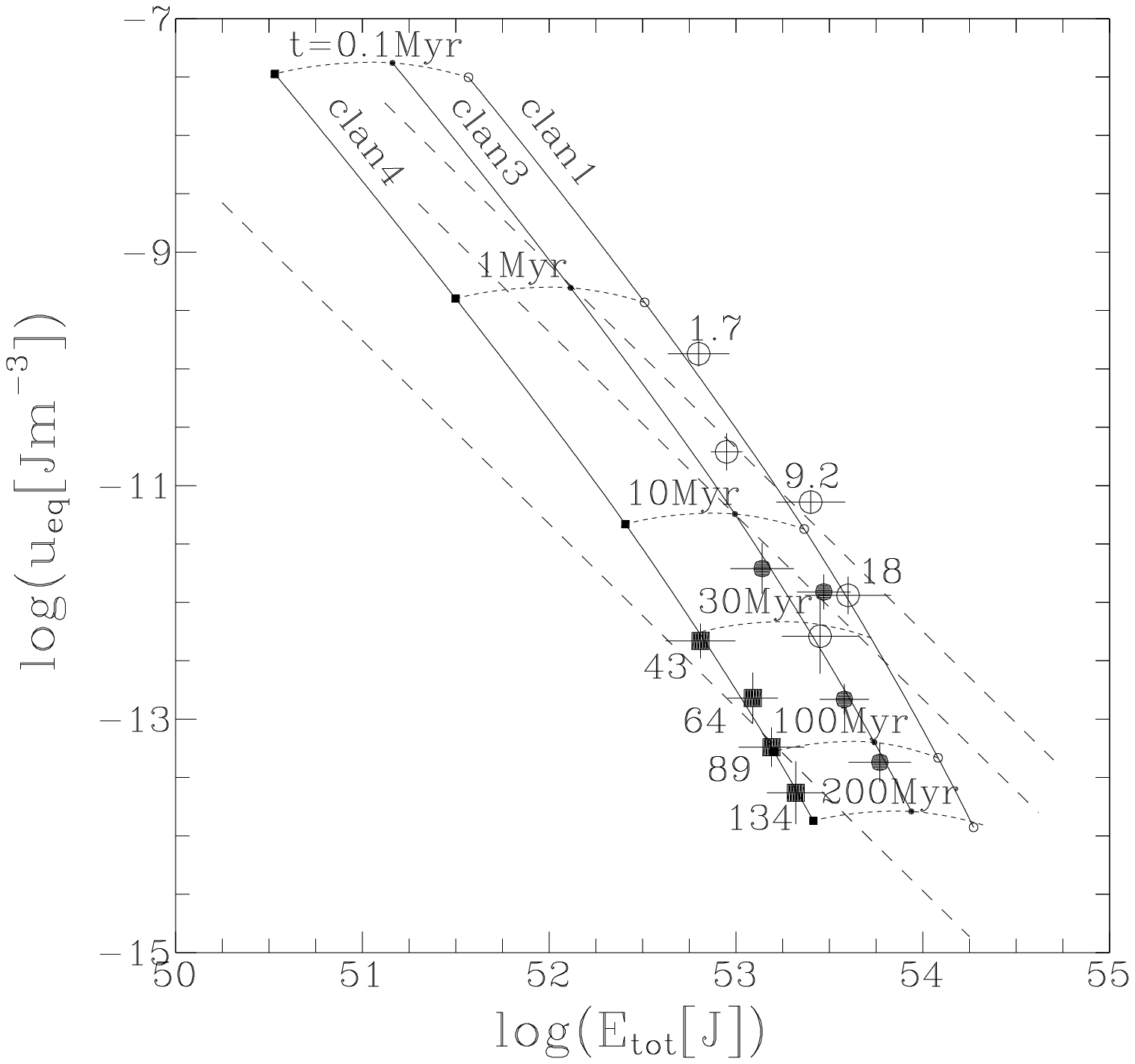}
\caption[]{Tracks of evolving energy fitted for the three clans. All the lines
and symbols are the same as in Figure 3.}
\end{figure}
  
As in Figure 3, the age markers are connected with dotted lines, and
the member sources of each clan are indicated with the same symbols. Note that
the ordinate of every source is its equipartition energy density derived from
the observational data, while the tracks show the {\sl model} cocoon
energy density. This is an additional argument that the modified model better
reproduce the observational data. In Figure~4, an uncertainty of the values of
both parameters is shown by the error bars. The error of $u_{\rm eq}$ is given
in Paper I (Table~2) and the error of $E_{\rm tot}$ is estimated from the error
of source volume (in Table~1 there). Again, the numbers mark the actual age of
the corresponding source. It is worth emphasizing that the modified evolutionary
$u_{\rm c}$--$E_{\rm tot}$ tracks are
steeper and curved in respect to those expected from the original model.
This effect, related to the rate of adiabatic losses and inflation of the
cocoon, is also discussed in Section 3.1.

\section{Discussion of the Results}

\subsection{Dynamical Evolution of Individual Sources}

The data show that the source (cocoon) geometry, described by its axial
ratio, depends on its estimated age.
In Sections 2.3 and 2.4 we have presented the evolutionary tracks of the `fiducial'
sources across the $P$--$D$ and $u_{\rm eq}$--$E_{\rm tot}$ planes, respectively,
calculated with incorporation of the cocoon axial ratio evolving with time
into the analytical KDA model. A predicted evolution of each
fiducial source, i.e. its tracks across the above planes, is verified by the
age, luminosity, size, and the energy of a clan members (cf. Section~2.2).
Therefore, we consider those sources as an individual source observed at
different lifetime epochs, and the relevant fiducial source is its
model representation.

Figures~3 and 4 show that for each of the three clans our tracks are steeper
than those resulting from the original KDA model, and much better corresponds to the
observed parameters of the clan sources. Our $P$--$D$ tracks strictly resemble
those published by Blundell et al., as both arise from the dynamical models without
assumption of the self-similar evolution of the source cocoon.
Also the tracks of our clans across the
$u_{\rm eq}$--$E_{\rm tot}$ plane are steeper than those predicted by the
original KDA model for sources with constant axial ratios. Moreover, the
steepening increases throughout the source lifetime. This is provided with the
time-dependent ${\cal P}_{\rm hc}$ ratio and non-constant inflation of the cocoon
applied in our analysis.
 
In the KDA model the synchrotron emitting
particles flow from the jet head into the lobe (cocoon) whose pressures are in a
constant ratio throughout the sources' lifetime. Thus the adiabatic losses in this
model do not increase with time. In the BRW model the population of emitting
particles adiabatically leave the constant-pressure hotspot area into the cocoon
whose pressure decreases throughout the sources' lifetime causing the losses due
to its adiabatic expansion increase with time. It is interesting that we get
similar tracks from a different approach, subsequently supporting predictions
of their model.

The observed distributions of the
clan sources across the $P$--$D$ and $u_{\rm eq}$--$E_{\rm tot}$ planes give
further arguments after the necessity of accounting for the evolving axial ratio
of the cocoon in studies of the dynamical evolution of individual radio sources,
and allow to conclude that a departure from the self similar expansion in
large and old sources is very probable.

All the analyzed clans include {\sl giant} sources, thus a dynamical
evolution of radio sources from typical (`normal') size to `giant' size is
in accordance with predictions of the analytical KDA and BRW models. This supports
our statistical findings that giants are old sources with high enough jet power
evolved in relatively low-density environment.

\subsection[section]{Deceleration/Acceleration of the Cocoon Expansion Speed}

The expansion speed of the cocoon along the jet axis can be directly derived
dividing length of the jet [equation (1) in Paper I] by time. Indicating
$g\equiv Q_{0}/\rho_{0}$ we have

\[{\cal V}_{\rm h}=L_{\rm j}/t\propto t^{(\beta -2)/(5-\beta)}g^{1/(5-\beta)}\]

\noindent
In the above relation, taken directly from the KDA model, an acceleration of the
cocoon expansion speed is admitted if $\beta>2$. The value assumed in Paper I
$\beta$=1.5 implies ${\cal V}_{h}\propto t^{-0.143}$. However this relation is
valid for the cocoon self-similar expansion only. As there is a doubt about validity
of this assumption for the large and old radio sources, the expansion speed can
rather be derived from the cocoon time-evolving volume of cylindrical geometry
given by

\begin{equation}
V_{\rm c}(t,g)=2\frac{\pi}{4R_{\rm T}^{2}}[L_{\rm j}(t,g)]^{3}\propto t^{9/(5-\beta)}
g^{3/(5-\beta)}.
\end{equation}

\noindent
(cf. equation (6) in Paper I).
Inserting $L_{\rm j}=D'/2$ and $R_{\rm T}=AR/2$ into Equation (1) and taking 1/3
root of it, we have
$D(t,g)\propto (AR)^{2/3}t^{3/(5-\beta)}g^{1/(5-\beta)}$. But the sample statistics
indicate that $AR$ also evolve with time and the ratio of $Q_{0}/\rho_{0}$.
Substituting

\[AR(t,g)\propto t^{0.16\pm 0.03}g^{0.082\pm 0.04}\]

\noindent
obtained from fitting a surface to the values of $AR$ over the $t$--$g$ plane
and $\beta=1.5$ into Equation (1), then dividing it by $t$, now we have

\begin{equation}
{\cal V_{\rm h}}(t,g)\propto t^{-0.04\pm 0.04}Q_{0}^{0.34\pm 0.03}.
\end{equation}

\noindent
The negative power exponent of $t$ still implies the cocoon expansion-speed
deceleration with time, though slower than that for the self-similarity expanding
cocoon.

Nevertheless, a slow {\em acceleration} effect is seen in the time evolution of the
{\sl fiducial} sources representing the clans of a few actual sample sources in
which each source in the clan with almost identical values of $Q_{0}$ and $\rho_{0}$
is observed at a different age (cf. Section~2.2 and Figure~2a). Below we analyse
whether it can be real.

Independent arguments after an accelerated expansion come from the
well known hydrodynamical considerations. From Scheuer (1974), it has commonly
been accepted that the ram pressure of the external gas behind the head of jet is

\begin{equation}
\rho_{\rm a}{\cal V}^{2}_{\rm h}\approx Q_{0}/(A_{\rm h}v_{\rm jet}),
\end{equation}

\noindent
where $\rho_{\rm a}$ is the external density,
$A_{\rm h}$ is the head working surface area, and the jet bulk velocity
$v_{\rm jet}$ is commonly assumed to be close to the light speed $c$. In this
approach ${\cal V}_{\rm h}$ is a function of the root square of the ratio
$Q_{0}/\rho_{\rm a}$ and the reciprocal of the linear size (diameter) of the
working area. $Q_{0}/\rho_{\rm a}$ simply represents an effectiveness of the jet
propagation across the surrounding medium. We can consider this ratio at the
core radius $a_{0}$ and at the source leading head where it is $Q_{0}/\rho_{0}$
and $Q_{0}/\rho_{\rm end}=Q_{0}/\rho_{0}(D/2a_{0})^{-\beta}$, respectively. Both
ratios are derived from the model. Values of the first ratio for {\sl giant}
sources are not very different from those for other sample sources being little
smaller than corresponding values for high-redshift sources, comparable with
those for low-redshift sources, and evidently higher from those for
low-luminosity ones. The second ratio, as a function of $D^{\beta}$, is the
highest for {\sl giant} sources.

The area  $A_{\rm h}$ at a given radius from the AGN centre can be determined
from high-resolution observations of the hotspots. For example, VLBI observations
of selected `compact symmetric objects' (CSO), possibly progenitors of classical
double FRI and/or FRII sources give a linear size of the working area of about
 a few parsecs at a radius of about 50$\div$100 pc (cf. Owsianik et al. 1998).
Many VLA observations of hotspots in sources of size 10$\div$20 kpc show
that their sizes are not larger than about 500 pc, while the hotspot sizes in
the lobes of large sources do not exceed  15 kpc. Schoenmakers et al.
(2000) in their study of {\sl giant} radio sources have assumed a typical
hotspot size of 5 kpc at a radius of $0.5\sim 1$ Mpc. After the discovery of the
third largest radio galaxy J1343+3758 (1343+379 in Tables~1 and 2), Machalski 
and Jamrozy (2000) determined a working surface area with a diameter of $\sim$13
kpc in one of its lobes.

From the data in our sample sources we have
$\langle Q_{0}/\rho_{0}\rangle =10^{61.9\pm 0.13}$
W\,m$^{3}$ kg$^{-1}$ and $\langle Q_{0}/\rho_{\rm end}\rangle =10^{64.9\pm 0.19}$
W\,m$^{3}$ kg$^{-1}$. Dividing the first ratio by $A_{\rm h}\approx (2\pm 0.7)\cdot10^{38}$
m$^{2}$ (the working area for a hotspot diameter of 0.5$\pm$0.1 kpc) and the second
ratio by $A_{\rm h}\approx (8\pm 6)\cdot10^{40}$ m$^{2}$ (for a hotspot diameter of 10$\pm$4
kpc), we
obtain ${\cal V}_{\rm h}$ of 0.12 and 0.19$c$, respectively, which confirms
a possible acceleration of the cocoon expansion speed. The above speeds, too high
with respect to an upper limit of ${\cal V}_{\rm h}$ (i.e. the quotient $D/2t$),
can be caused by either an overestimation of the jet power or too low internal
density of the core and of the external density at the end of jet.
An adjustment of $\beta$ will not work because some increase of
$\beta$ can compensate for too low $\rho_{0}$, but will dramatically lower
$\rho_{\rm end}$. Since the values of $A_{\rm h}$ adopted for this calculation
may be overestimated rather than underestimated, only a decrease of $Q_{0}$ can
fit the observed expansion speeds.

Indeed, the values of $Q_{0}$
derived from equation (3) base on the ram pressure considerations in the
overpressured source model A of Scheuer (1974) and its further modifications
(e.g. Begelman and Cioffi 1989; Loken et al. 1992; Nath 1995). They all are
self-similar models of the Carvaldo and O'Dea (2002) type I which describe the source
dynamics only. If the source energetics and especially the energy losses are
taken into account (the type III models; e.g. KDA, BRW), the significantly higher
values of $Q_{0}$ are implied. Therefore, the above calculation of 
${\cal V}_{\rm h}$ values must be based on $Q_{0}$ values determined within the
same dynamical model of type I. If this is the case, the controversy about too
high expansion speeds will disappear. Using the Nath's (1995) model in which 100\%
of the twin-jets energy is converted into radiation (i.e. $2Q_{0}t=U_{\rm eq}$),
we find that $Q_{0}$ values fitted with that model are about ten times lower than
the KDA values. Therefore, the ${\cal V}_{\rm h}$ values calculated as above but
with the $Q_{0}$(NATH) values decrease to about 0.04 and 0.06$c$, i.e. to the
expansion velocities quite plausible. 

\section{Conclusions}

The following conclusions can be drawn from the above analysis:

(1) The sample sources, having similar model parameters $Q_{0}$ and $\rho_{0}$
but different age, size, axial ratio and luminosity (called a `clan'), fit
satisfactory the modified Kaiser et al.'s (1997) [KDA] model tracks on the $P-D$
and $u_{\rm eq}-E_{\rm tot}$ planes
calculated for a `fiducial' source being the model representation of each clan.
The derived tracks appear to be much steeper than those provided by the original
KDA model; the best fit is achieved by allowing an evolution of the sources' cocoon
axial ratio with time. The derived tracks are
compatible with those expected from the more sophisticated Blundell et al.'s
(1999) model.

(2) The evolution of the energetics of sources on the $u_{\rm eq}-E_{\rm tot}$, 
predictable from the model, form another characteristic of their population 
which can be constrained by observational data.

(3) In the case of clan sources, we found a slow acceleration of the average 
expansion speed of the cocoon along the jet axis. This effect is predicted with 
the model modified as above, and is caused by a systematic increase of
the ratio between hotspot and cocoon pressures (${\cal P}_{\rm hc}$) with the
age of source and decreasing density of the external environment according to the
King's law model. The statistics suggest that this acceleration is also dependent
on an efficiency of the jet propagation through the surrounding medium.


\begin{thebibliography}{}

\bibitem{begelman}
{Begelman, M.C., and Cioffi, D.F.} {1989,} {\it ApJ,} {345,} {L21}
\bibitem{blundell}
{Blundell, K.M., Rawlings, S., and Willott, C.J.} {1999,} {\it AJ,} {117,} {766}
\bibitem{carvalho}
{Carvalho, J.C., and O'Dea, C.P.} {2002,} {\it ApJS,} {141,} {337}
\bibitem{falle}
{Falle, S.A.E.G.} {1991,} {\it MNRAS,} {250,} {581}
\bibitem{fanaroff}
{Fanaroff, B.L., and Riley, J.M.} {1974,} {\it MNRAS,} {167,} {31P}
\bibitem{gopal}
{Gopal-Krishna and Wiita, P.J.} {2001,} {\it ApJ,} {560,} {L115}
\bibitem{hardcastle}
{Hardcastle, M.J., and Worrall, D.M.} {2000,} {\it MNRAS,} {319,} {562}
\bibitem{kaiser1}
{Kaiser, C.R., Dennett-Thorpe, J., and Alexander, P. (KDA)} {1997,} {\it MNRAS,} {292,} {723}
\bibitem{kaiser2}
{Kaiser, C.R.} {2000,} {\it A\&A,} {362,} {447}
\bibitem{loken}
{Loken, C., Burns, J.O., Clarke, D.A., and Norman, M.L.} {1992,} {\it ApJ,} {392,} {54}
\bibitem{mach1}
{Machalski, J., and Jamrozy, M.} {2000,} {\it A\&A,} {363,} {L17}
\bibitem{mach2}
{Machalski, J., Chy\.{z}y, K.T., and Jamrozy, M.} {2004,} {\it Acta Astr.,} {54,} {249}
\bibitem{nath}
{Nath, B.B.} {1995,} {\it MNRAS,} {274,} {208}
\bibitem{owsianik}
{Owsianik, I., Conway, J.E., and Polatidis, A.G.} {1998,} {\it A\&A,} {336,} {L37}
\bibitem{rawlings}
{Rawlings, S., and Saunders, R.} {1991,} {\it Nature,} {349,} {138}
\bibitem{scheuer}
{Scheuer, P.A.G.} {1974,} {\it MNRAS,} {166,} {513}
\bibitem{schoen3}
{Schoenmakers, A,P., Mack, K.-H., de Bruyn, A.G., R\"{o}ttgering, H.J.A., 
Klein,~U., and van der Laan, H.} {2000,} {\it A\&AS} {146,} {293}

\end{thebibliography}
\end{document}